\documentclass[conference,letterpaper]{IEEEtran}

\usepackage[utf8]{inputenc} 
\usepackage[T1]{fontenc}
\usepackage{url}
\usepackage{ifthen}
\usepackage{cite}
\usepackage[cmex10]{amsmath} 

\usepackage{cite}
\usepackage{amssymb,amsfonts}
\usepackage{graphicx}
\usepackage{textcomp}
\usepackage{xcolor}
\usepackage{comment}
\usepackage{algorithm}
\usepackage{algpseudocode}
\usepackage{float}
\usepackage{bbm}
\usepackage{algorithm}
\usepackage{algpseudocode}

\usepackage{tikz,pgf}
\usetikzlibrary{patterns}
\usetikzlibrary{calc}

\usepackage{pgfplots}

\newtheorem{definition}{Definition}

\def\BibTeX{{\rm B\kern-.05em{\sc i\kern-.025em b}\kern-.08em
    T\kern-.1667em\lower.7ex\hbox{E}\kern-.125emX}}

\IEEEoverridecommandlockouts

\begin{document}

\title{Optimal Resource Allocation for Cellular \\ Networks with Virtual Cell Joint Decoding
\thanks{This research was supported by Huawei, by AFOSR Grant FA9550-12-1-0215, and by ONR Grants N00014-15-1-2527 and N00014-18-1-2191.}
}

\author{\IEEEauthorblockN{ Michal Yemini}
\IEEEauthorblockA{Wireless Systems Laboratory\\
Stanford University\\
California, USA \\
michalye@stanford.edu}
\and
\IEEEauthorblockN{Andrea J. Goldsmith}
\IEEEauthorblockA{Wireless Systems Laboratory\\
Stanford University\\
California, USA \\
andrea@wsl.stanford.edu}

}

\maketitle

\begin{abstract}
This work presents a new resource allocation optimization framework for cellular networks  using  neighborhood-based optimization. Under this optimization framework resources are allocated within virtual cells encompassing several base-stations and the users within their coverage area. Incorporating the virtual cell concept enables the utilization of more sophisticated cooperative communication schemes such as coordinated multi-point decoding. We form the virtual cells using hierarchical clustering  given a particular number of such cells. Once the virtual cells are formed, we consider a cooperative decoding scheme in which the base-stations in each virtual cell jointly decode the signals that they receive. We propose an iterative solution for the resource allocation problem  resulting from the cooperative decoding within each virtual cell.  Numerical results for the average system sum rate of our network design under hierarchical clustering are presented. These results indicate that virtual cells with neighborhood-based optimization leads to significant gains in sum rate over optimization within each cell, yet may also have a significant sum-rate penalty compared to fully-centralized optimization. 
\end{abstract}


\section{Introduction}
 The  increased capacity demand in cellular networks  is a major driver in the deployment of 5G systems. To increase network capacity, the deployment of small cells has been proposed and is currently taking place  \cite{4623708,6768783,6171992,anpalagan_bennis_vannithamby_2015}. The main caveat of the usage of small cells is that their proximity to one another combined with their frequency reuse can cause severe interference which must be managed carefully to maximize the overall network capacity.    
 To reduce interference a new interference mitigation paradigm  called Cooperative Multi-Point (CoMP) was proposed (see \cite{5706317}).  This paradigm encompasses several cooperation models such as Uplink Interference Prediction in which cooperation is allowed in the resource allocation stage only, and the Uplink Joint Detection model that we consider in this work which allows base-station cooperation in both the resource allocation and decoding stages. We investigate a flexible cooperative resource allocation structure for cellular systems  where, instead of each base-station serving all users within its own cell independently, several base-stations act cooperatively to create a “virtual cell” in which the base-stations jointly decode their signals.
To  design wireless networks that are composed of virtual cells we address in this work the following two design challenges: 1) Creating the virtual cells, i.e., clustering the base-stations into virtual cells. 2) Allocating the resources in each virtual cell assuming cooperative decoding with infinite-capacity backhaul links between base-stations in each virtual cell.

Base-station and user clustering as part of network performance enhancement  is discussed in the CoMP literature, see for example  \cite{7839266,6530435,6707857,6181826,6555174,5594575,5502468,6655533,4533793,5285181,6786390,8260866}.   The clustering of base-stations and users can be divided into three groups: 1) Static clustering which considers a cellular network whose cells are clustered statically. Hence, the clustering does not adapt to network changes. Examples for static clustering algorithms are presented in \cite{6530435,6181826,6707857,6555174}. 2) Semi-dynamic clustering, in which static clusters are formed but the cluster affiliation of users is adapted according to the networks changes. Examples for such algorithms are presented in  \cite{5594575,5502468,6655533}. 3) Dynamic clustering in which the clustering of both base-stations and users adapts to changes in the network. Examples for dynamic clustering algorithms are presented in \cite{4533793,5285181,6786390}. For an extensive literature survey of cell clustering for CoMP in wireless networks see the work \cite{7839266}. Finally, cell clustering strategies in wireless networks is also investigated in the ultra-dense network literature, see for example \cite{7008373,Tang2015,7498053,7794900,8284755,5963458}.

Resource allocation for virtual cell joint decoding is closely related to cloud radio access network  \cite{5594708,CIT-048,6924850,7487951,6601765} in which several cells act cooperatively.   Interestingly, maximizing the user sum rate in a virtual cell is equivalent to maximizing the sum rate of a multiple access channel with  multiple receiving antennas and several frequency bands. Thus, the optimal resource allocation scheme in a virtual cell in terms of user sum rate is capacity-achieving. Furthermore, this optimal resource allocation can be calculated by convex optimization techniques.  

\section{Problem Formulation}\label{sec:problem_formualtion}
We consider  a communication network that comprises a set of base-stations (BSs) $\mathcal{B}$, a set of users $\mathcal{U}$ and a set of frequency bands $\mathcal{K}$. The users communicate with the BSs which can choose to cooperatively decode their signals. Each user $u\in\mathcal{U}$ has a power constraint of $\overline{P}_u$ dBm.
To form the  neighborhood in which decoding is performed cooperatively the BSs and users are clustered into virtual cells which must fulfill the following characteristics.

\subsection{Virtual Cells}\label{sec:virtual_cell_requirements}

\begin{definition}[Virtual BS]
Let $b_1,..,b_n$ be $n$ BSs in a communication network, we call the set $\{b_1,..,b_n\}$ a virtual BS.
\end{definition}
\begin{definition}[Proper clustering]	
Let $\mathcal{B}$ be a set of BSs,  $\mathcal{U}$ be a set of users. Denote   $\mathcal{V}=\{1,\ldots,V\}$.
For every $v$, define the sets $\mathcal{B}_v\subset \mathcal{B}$ and $\mathcal{U}_v\subset \mathcal{U}$ .
We say that the set $\mathcal{V}$ is a proper clustering of the sets  $\mathcal{B}$ and  $\mathcal{U}$  if $\mathcal{B}_v$ and $\mathcal{U}_v$  are partitions of the sets $\mathcal{B}$ and $\mathcal{U}$, respectively. That is,
$\bigcup_{v\in\mathcal{V}}\mathcal{B}_v = \mathcal{B}$, $\bigcup_{v\in\mathcal{U}}\mathcal{U}_v = \mathcal{U}$. Additionally,
  $\mathcal{B}_{v_1}\cap\mathcal{B}_{v_2}=\emptyset$ and $\mathcal{U}_{v_1}\cap\mathcal{U}_{v_2}=\emptyset$ for all $v_1,v_2\in\mathcal{V}$ such that $v_1\neq v_2$.
\end{definition}

\begin{definition}[Virtual cell]
Let $\mathcal{B}$ be a set of BSs, $\mathcal{U}$ be a set of users, and  $\mathcal{V}$ be a proper clustering of $\mathcal{B}$ and $\mathcal{U}$. For every $v\in\mathcal{V}$  the virtual cell  $\mathcal{C}_v$ is composed of the virtual BS $\mathcal{B}_v$ and the set of users $\mathcal{U}_v$.	
\end{definition}

This condition ensures that every BS and every user belongs to exactly one virtual cell.

Let $\mathcal{V}$ be a proper clustering of the set of BSs $\mathcal{B}$ and the set of users $\mathcal{U}$, and let   $\{\mathcal{C}_v\}_{v\in\mathcal{V}}$ be the set of virtual cells that $\mathcal{V}$ creates.
In each virtual $\mathcal{C}_v$ we assume that the BSs that compose the virtual BS $\mathcal{B}_v$ allocate their resources jointly.

\subsection{The Uplink Resource Allocation Problem for CoMP Decoding}\label{subsection:uplink_joint_decoding_problem}

In each virtual cell we consider uplink cloud decoding with infinite capacity. For a single frequency band scenario, this setup is equivalent to a multiple access channel  with multiple users, each with a single transmitting antenna, and  multiple receiving antennas. Denote by $x_{u,k}$ the signal of user $u$ in frequency band $k$, and by $y_{b,k}$  the received signal at BS $b$ in frequency band $k\in\mathcal{K}$. For the sake of clarity, we label the BSs in the cluster $v$ by $b_1,\ldots, b_{|\mathcal{B}_v|}$, and label the users in cluster $v$ by  $u_1,\ldots,u_{|\mathcal{U}_v|}$.  
Denote  $\boldsymbol y_{v,k}\triangleq(y_{b_1,k},\ldots,y_{b_{|\mathcal{B}_v|},k})^T$ and let $\boldsymbol x_{v,k}\triangleq(x_{u_1,k},\ldots,x_{u_{|\mathcal{U}_v|},k})^T$, where $(\cdot)^T$ denotes the transpose operator. The receiving signal at BS $b\in \mathcal{B}_v$, ignoring the interference from other virtual cells, in frequency band $k$ is
$
y_{b,k} = \sum_{i=1}^{|\mathcal{U}_v|}h_{u_i,b,k} x_{u_i,k}+n_{b,k}
$,
where $h_{u_i,b,k}$ is the channel coefficient from user $u_i$ in $v$ to the BS $b$ in $v$ over frequency band $k$, and $n_{b,k}$ is a white Gaussian noise at BS $b$ over frequency band $k$. 

Let $\boldsymbol h_{u_i,k} = (h_{u_1,b_1,k},\ldots,h_{u_i,b_{|\mathcal{B}_v|},k})'$ be the channel coefficient vector between user $u_i$ in $v$ to all the BSs in cluster $v$, then the receiving signal vectors at the BSs in $v$ is
$
\boldsymbol y_{v,k} = \sum_{i=1}^{|\mathcal{U}_v|}\boldsymbol h_{u_i,k} x_{u_i,k}+\boldsymbol n_{v,k}
$,
where 
$\boldsymbol n_{v,k}=(n_{b_1,k},\ldots,n_{b_{|\mathcal{B}_v|,k}})$ is a white noise vector at the BSs. 
Denote $\boldsymbol C_{v,k}=\text{cov}\left(\boldsymbol{x}_{v,k}\right)$ and $\boldsymbol N_{v,k} = \text{cov}(\boldsymbol n_{v,k})$ and let $W_k$ be the bandwidth of frequency band $k$; the sum capacity of the uplink, ignoring interference outside the virtual cell,  in the virtual cell is then:
\begin{flalign}\label{eq:uplink_problem_clean}
\max &\sum_{k\in\mathcal{K}}W_k\log_2\left|\boldsymbol I+\sum_{u\in\mathcal{U}_v}p_{u,k}\boldsymbol h_{u,k} \boldsymbol h_{u,k}^{\dagger}\boldsymbol{N}_{v,k}^{-1}\right|\nonumber\\
\text{s.t.: } & \sum_{k\in\mathcal{K}} p_{u,k}\leq \overline{P}_u, \quad p_{u,k}\geq 0 \:\:\: \forall k\in\mathcal{K},
\end{flalign}
where  $\boldsymbol h_{u,k}^{\dagger}$ is the conjugate transpose of  $\boldsymbol h_{u,k}$ and $|\boldsymbol A|$ denotes the determinant of the matrix $\boldsymbol A$.  We assume that the matrix $\boldsymbol N_{v,k}$ is invertible for all $k$ and thus also positive definite.

We note that the problem (\ref{eq:uplink_problem_clean}) ignores interference that is caused by transmissions outside the virtual cell, instead it only considers interference that is caused by transmissions in the virtual cell. However, as   the number of virtual cells decreases, this interference  becomes the dominant one. Indeed, numerical results show that incorporating virtual cells improves performance monotonically as the number of virtual cells in the network decreases.   

\section{Forming the Virtual Cells}\label{sec:virtual_cell_create}
This section presents the clustering approaches that create the virtual cells within which the resource allocation scheme we present in Section \ref{sec:joint_decoding_resource} operates.
We consider two methods to cluster the BSs. The first is a hierarchical clustering of the BS according to a minimax linkage criteria. To evaluate the performance of this clustering method we compare it  to an exhaustive search over all the possible clusterings of BSs.
\subsection{Base-Station Clustering}\label{sec:BS_clustering}
\paragraph{Hierarchical clustering - Minimax linkage \cite{BienTibshirani2011}}
Let $d:\mathbb{R}^2\times\mathbb{R}^2\rightarrow\mathbb{R}$ be the Euclidean distance function.
\begin{definition}[Radius of a set around point]
	Let $S$ be a set of points in $\mathbb{R}^2$, the radius of $S$ around $s_i \in S$  is defined as $r(s_i,S)=\max_{s_j\in S}\:d(s_i,s_j)$.
\end{definition}
\begin{definition}[Minimax radius]
	Let $S$ be a set of points in $\mathbb{R}^2$, the minimax radius of $S$ is defined as $r(S) = \min_{s_i\in S}\: r(s_i,S)$.
\end{definition}
\begin{definition}[Minimax linkage]
	The minimax linkage between two sets of points $S_1$ and $S_2$ in $\mathbb{R}^l$ is defined as $d(S_1,S_2) = r(S_1\cup S_2)$.
	
	Note that $d(\{s_1\},\{s_2\}) = r(\{s_1\}\cup \{s_2\})=d(s_1,s_2)$.
\end{definition}
 
Algorithm \ref{algo:hierarchical_clustering} presents the  hierarchical clustering algorithm using the minimax linkage criterion; it gets  a set of points $S$ and produces the clusterings $B_1,\ldots,B_n$, where $B_m$ is the clustering of size $m$.
Let $S=\{s_1,\ldots,s_n\}$ be the set of locations of the BSs in $\mathcal{B}$. We use Algorithm \ref{algo:hierarchical_clustering} with the input $S$ to create the  virtual BSs for each number of clusters $m$.
\begin{algorithm}
	\caption{}\label{algo:hierarchical_clustering}
	\begin{algorithmic}[1]

\State Input: $S=\{s_1,\ldots,s_n\}$;
\State Set $B_n = \left\{\{s_1\},\dots,\{s_n\}\right\}$; 
\State Set $d(\{s_i\},\{s_j\})=d(s_i,s_j),\:\forall s_i,s_j\in S$;

\For {$m = n-1,\ldots,1$}
	\State Find $(S_1,S_2) = \arg\min_{G,H\in B_m: G\neq H} d(G,H)$;
	\State  Update $B_{m} = B_{m+1} \bigcup \{S_1\cup S_2\} \setminus \{S_1,S_2\}$;
	\State Calculate $d(S_1\cup S_2,G)$ for all $G\in B_m$;
\EndFor
\end{algorithmic}

\end{algorithm}

The hierarchical clustering is important to our setup since it enjoys a key
 property that both the K-means clustering and the spectral clustering lack, namely, the number of clusters can be changed without disassembling all the clusters in the networks. Thus, the  number of virtual BSs can be easily adapted according to the current state of the network. Moreover, at each stage of Algorithm \ref{algo:hierarchical_clustering} the minimax linkage criterion minimizes the radius of the new cluster that is created by the merging of two existing  clusters. Since interference increases on average as distance decreases this criterion  merges  two clusters to create a new one in which the minimal interference is maximized on average; this interference is then optimized in the resource allocation scheme.  Finally, the minimax linkage criterion enjoys several desirable  properties that are discussed in \cite{BienTibshirani2011}. 

\paragraph{Exhaustive Search} To evaluate  the performance of hierarchical clustering  against a theoretical upper bound we  also performed exhaustive search over all the possible clusterings of BSs. In this way, for a given number of clusters (virtual BSs) we produced all the possible clusterings of BSs and in the end, after calculating the power allocation of all the virtual cells, chose the clustering that yielded the maximal sum rate of the network. This maximal sum rate considered interference from other virtual cells, given the number of clusters and the user affiliation rule.

\subsection{Users' Affiliation with Clusters}\label{sec_user_affil}
To create the virtual cells, we consider two affiliation rules: 1) Closest BS rule in which each user is affiliated with its closest BS. 2) Best channel rule in which  each user is affiliated with the BS to which it has the best channel  (absolute value of the channel coefficient). Then each user is associated with the virtual BS that its affiliated BS is part of.
This way every virtual BS and it associated users compose a virtual cell.

It is easy to verify that these formations of the virtual cells fulfill the requirement presented in Section \ref{sec:virtual_cell_requirements}.

\section{Resource Allocation for Joint Decoding}\label{sec:joint_decoding_resource}
This section is dedicated to solving the problem (\ref{eq:uplink_problem_clean}) that is presented in Section \ref{subsection:uplink_joint_decoding_problem}
in which BSs use cloud decoding with backhaul links of infinite capacity.  

Using the identity $|\boldsymbol{AB}|=|\boldsymbol{A}|\cdot|\boldsymbol{B}|$ we have that the capacity of the virtual cell is
\begin{flalign}\label{problem_joint_decode_ininite}
\max &\sum_{k\in\mathcal{K}}W_k\left[\log_2\left|\boldsymbol{N}_{v,k}+\sum_{u\in\mathcal{U}_v}p_{u,k}\boldsymbol h_{u,k} \boldsymbol h_{u,k}^{\dagger}\right|-\log_2\left|\boldsymbol{N}_{v,k}\right|\right]\nonumber\\
\text{s.t.: } & \sum_{k\in\mathcal{K}} p_{u,k}\leq \overline{P}_{u,k},\quad p_{u,k}\geq 0.
\end{flalign}
Since the terms $\log_2\left|\boldsymbol{N}_{v,k}\right|$ are constants, hereafter we omit them from the objective function.
Denote $\boldsymbol{p}_u = (p_{u,1},\ldots,p_{u,K})$ and let:
\begin{flalign*}
&f\left(\boldsymbol{p}_{u_1},\ldots,\boldsymbol{p}_{u_{|\mathcal{U}_v|}}\right) = \log_2\left|\boldsymbol{N}_{v,k}+\sum_{u\in\mathcal{U}_v}p_{u,k}\boldsymbol h_{u,k} \boldsymbol h_{u,k}^{\dagger}\right|.
\end{flalign*} 

The following three conditions must hold in order to optimally solve the problem (\ref{problem_joint_decode_ininite}) iteratively using the cyclic  coordinate ascend algorithm \cite[Chapter 2.7]{Bertsekas/99}:
\begin{enumerate}
	\item The function $f\left(\boldsymbol{p}_{u_1},\ldots,\boldsymbol{p}_{u_{|\mathcal{U}_v|}}\right)$ is concave.
	\item Define 
	\begin{flalign}
	\mathcal{P}&\triangleq  \left\{\left(\boldsymbol{p}_{u_1},\ldots,\boldsymbol{p}_{u_{|\mathcal{U}_v|}}\right):\sum_{k\in\mathcal{K}} p_{u,k}\leq \overline{P}_u,\right.\nonumber\\
	&\hspace{3.25cm}\left.\: \sum_{k\in\mathcal{K}}p_{u,k}\geq 0 \:\: \forall\: u\in\mathcal{U}_v\right\},\nonumber\\
	\mathcal{P}_u&\triangleq\left\{\boldsymbol{p}_u:\sum_{k\in\mathcal{K}}p_{u,k}\leq \overline{P}_u,\:p_{u,k}\geq0\right\},
	\end{flalign} 
	then $\mathcal{P} = \mathcal{P}_{u_1}\times\ldots\times\mathcal{P}_{u_{|U|}}$.
	\item The problem
	\begin{flalign}\label{problem_joint_decode_ininite_single}
	\max_{\tilde{\boldsymbol{p}}_{u_i}}\: &f\left(\boldsymbol{p}_{u_1},\ldots,\boldsymbol{p}_{u_{i-1}},\tilde{\boldsymbol{p}}_{u_i},\boldsymbol{p}_{u_{i+1}},\boldsymbol{p}_{u_{|U|}}\right)\nonumber\\
	\text{s.t.: } & \tilde{\boldsymbol{p}}_{u_i}\in\mathcal{P}_{u_i},
	\end{flalign}
has a unique maximizing solution.
\end{enumerate}

Next we solve the problem (\ref{problem_joint_decode_ininite_single}) and show the  optimal solution is uniquely attained.
Denote $\boldsymbol\Sigma_{i,k} = \boldsymbol{N}_{v,k}+\sum_{\substack{j\neq i,\\ u_j\in\mathcal{U}_v}}p_{u_j,k}\boldsymbol h_{u_j,k}\boldsymbol h_{u_j,k}^{\dagger}$. The problem  (\ref{problem_joint_decode_ininite_single})
is then
\begin{flalign}\label{problem_joint_decode_ininite_single_eq}
\max &\sum_{k\in\mathcal{K}}W_k\log_2\left|\boldsymbol\Sigma_{i,k}+p_{u_i,k}\boldsymbol h_{u_i,k} \boldsymbol h_{u_i,k}^{\dagger}\right|\nonumber\\
\text{s.t.: } & \sum_{k\in\mathcal{K}} p_{u_i,k}\leq \overline{P}_{u_i},\quad p_{u_i,k}\geq 0 \:\:\:\forall k\in\mathcal{K}.
\end{flalign}

The Karush–Kuhn–Tucker (KKT) conditions for (\ref{problem_joint_decode_ininite_single_eq}) are
\begin{flalign}
& W_k\frac{\boldsymbol h_{u_i,k}^{\dagger}\boldsymbol\Sigma_{i,k}^{-1}\boldsymbol h_{u_i,k}}{1+\boldsymbol h_{u_i,k}^{\dagger}\boldsymbol\Sigma_{i,k}^{-1}\boldsymbol h_{u_i,k}p_{u_i,k}}-\lambda+\mu_k = 0,\nonumber\\
& \lambda\left(\sum_{k\in\mathcal{K}} p_{u_i,k}-\overline{P}_{u_i}\right) = 0,\quad  \mu_kp_{u_i,k}=0,\nonumber \\
& \sum_{k\in\mathcal{K}}p_{u_i,k}\leq\overline{P}_{u_i},\quad p_{u_i,k}\geq 0,\nonumber\\
&\mu_k\geq 0, \quad \lambda\geq0 \:\:\:\forall k\in\mathcal{K}.
\end{flalign}

Since $\mu_k$ is nonnegative for all $k$, and the matrix $\boldsymbol\Sigma^{-1}_{i,k}$ is positive definite\footnote{Since $\boldsymbol{N}_{v,k}$ is a positive definite matrix, $\boldsymbol\Sigma_{i,k}$ is positive definite as well.} for all $k$, in order to fulfill 
the first KKT condition $\lambda$ must be strictly positive.  
Now, if $p_{u_i,k}>0$, then $\mu_k =0 $ and by the first KKT condition 
\begin{flalign}
p_{u_i,k}=\frac{W_k}{\lambda}-\frac{1}{\boldsymbol h_{u_i,k}^{\dagger}\boldsymbol\Sigma_{i,k}^{-1}\boldsymbol h_{u_i,k}}.
\end{flalign} 
Also, if $p_{u_i,k}=0$, then by the first KKT condition
$
W_kh_{u_i,k}^{\dagger}\boldsymbol\Sigma_{i,k}^{-1}\boldsymbol h_{u_i,k}+\mu_k = \lambda.
$
It follows that 
\begin{flalign}
p_{u_i,k} =  \left(\frac{W_k}{\lambda}-\frac{1}{\boldsymbol h_{u_i,k}^{\dagger}\boldsymbol\Sigma_{i,k}^{-1}\boldsymbol h_{u_i,k}}\right)^+
\end{flalign}
where $\lambda$ is chosen such that $\sum_{k\in\mathcal{K}}p_{u_i,k} = \overline{P}_{u_i}$.

\section{Numerical Results}\label{se:simulation}
This section presents Monte Carlo simulation results that compare the resource allocation and user affiliation schemes for both the hierarchical clustering and the exhaustive search over all possible clustering. We set the following parameters for the simulation: the network is comprised of $6$ BSs and $50$ users which were uniformly located in a square of side $2000$ meters.  There were $10$ frequency bands each of bandwidth 20 KHz, the carrier frequency was set to $1800$ MHz. The noise power received by each BS was $-174$ dBm/Hz, and the maximal power constraint for each user was $23$ dBm. Finally, in each frequency band we consider Rayleigh fading, Log-Normal shadowing with standard deviation of $8$ dB and a path loss model of $PL(d)= 35\log_{10}(d)+34$ where $d$ denotes the distance between the transmitter and the receiver in meters (see \cite{4138008}).
We averaged the results over 500 realizations, in each we generated randomly the locations of the BSs, users and channel coefficients.
In the simulation we compared the average sum rate achieved by each of the clustering and user affiliation schemes presented in this paper for cooperative decoding. We note that while the resource allocation ignored the interference caused by other virtual cells, the sum rate of the network was calculated considering this interference in the SINR of each user and the corresponding sum rate of that user.
\begin{figure}
	\centering
	\includegraphics[scale=0.65]{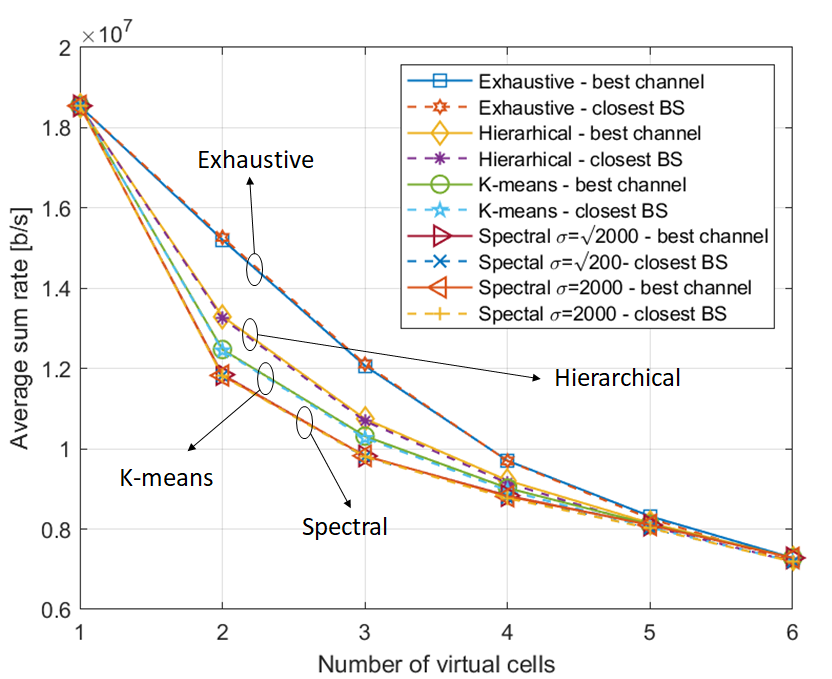}
	\caption{Average sum rate as a function of the number of virtual cells. The legend is written in the form X-Y where X and Y indicate the BS clustering algorithm and the user affiliation rule, respectively.}
	\label{Joint_sum_rate_exhaust_fig}
	\vspace{-0.35cm}
\end{figure}

Fig.~\ref{Joint_sum_rate_exhaust_fig} compares the average system sum rate as a function of the number of virtual cells for the resource allocation scheme presented in Section   \ref{sec:joint_decoding_resource}.  We compared the performance of the two BS clustering methods presented in \ref{sec:BS_clustering} with those of other BS clustering methods, namely K-means and spectral clustering. 
Fig.~\ref{Joint_sum_rate_exhaust_fig} leads to several interesting insights and conclusions. First, it confirms the expectation that as the number of virtual cells decreases, the average sum rate increases. Second, it shows that for a sufficiently large number of frequency bands, the  closest BS user affiliation rule and the best channel affiliation rule lead to similar performance. Third, it compares the performance of the hierarchical clustering and the exhaustive search over all BS clustering. This comparison illustrates that, while the exhaustive search outperforms the hierarchical clustering as expected, hierarchical clustering has similar performance with a much lower complexity. Additionally, Fig.~\ref{Joint_sum_rate_exhaust_fig}  shows that clustering BSs using the hierarchical approach with the minimax linkage criterion outperforms clustering the BSs  using either K-means or spectral clustering \cite{Ng:2001:SCA:2980539.2980649} algorithms, where the spectral clustering was performed for two possible values of $\sigma$: $\sqrt{2000}$ and $2000$. Both of these values yielded similar network performance.  

\begin{figure}
	\centering
	\includegraphics[scale=0.65]{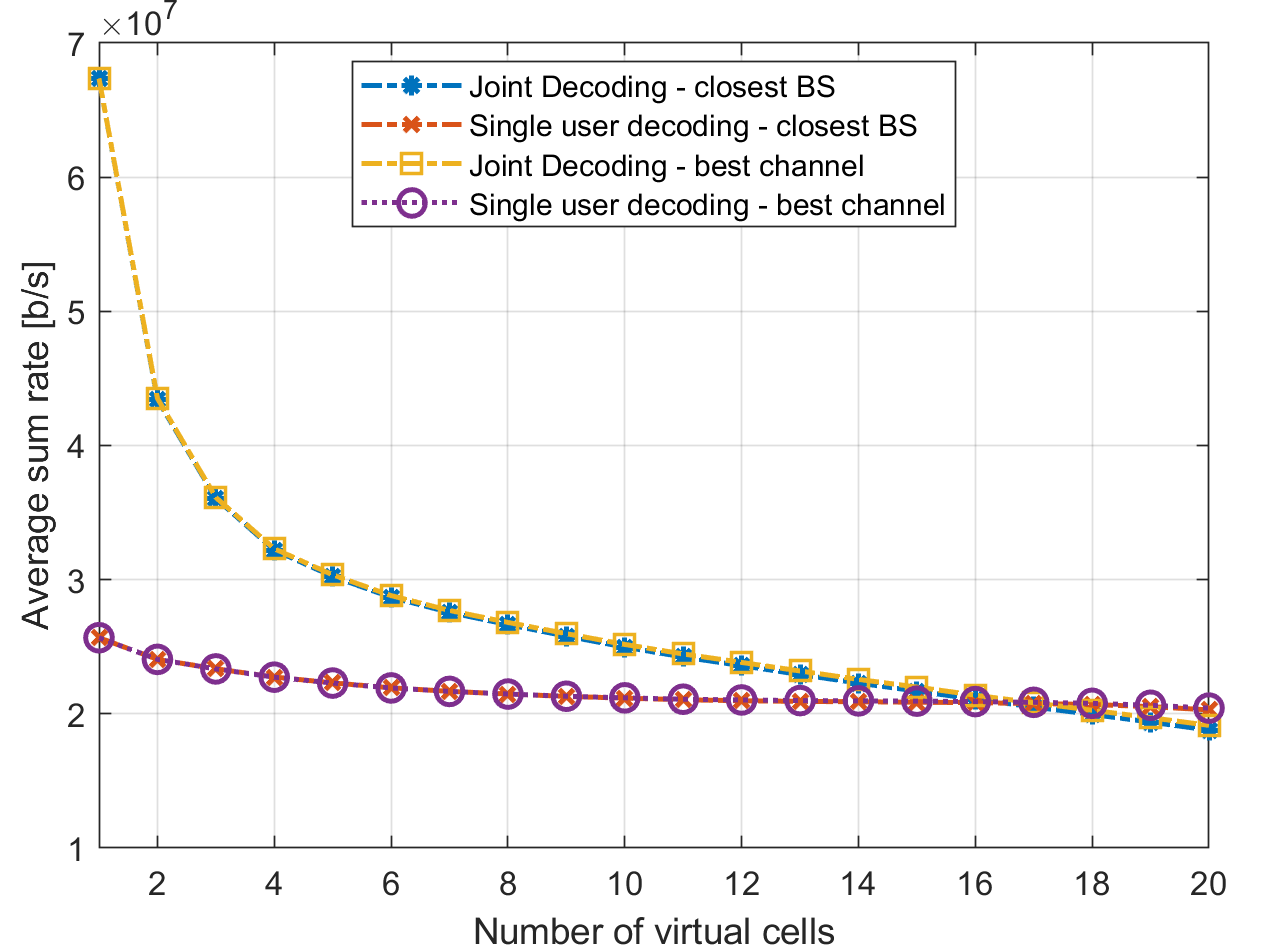}
	\caption{Comparison between the average sum rate  of joint and single user decoding as a function of the number of virtual cells using hierarchical BS clustering with minimax linkage criterion.}
	\label{Joint_sum_rate_hierarhical_fig}
	\vspace{-0.35cm}
\end{figure}

We also compared the average system sum rate achieved by joint decoding to the one achieved by single user decoding. The resource allocation problem for single user decoding is nonconvex, so there are multiple methods to approximately solve it, as detailed in \cite{YeminiGoldsmith1}. Fig.~\ref{Joint_sum_rate_hierarhical_fig} compares the average system sum rate achieved by joint decoding against the maximal average system sum rate achieved by the single user decoding methods presented in \cite{YeminiGoldsmith1}, for each number of virtual cells. The virtual cells were generated by using the hierarchical clustering presented in Algorithm \ref{algo:hierarchical_clustering}. We simulated a network with a larger number of users and BSs, specifically $100$ users and $20$ BSs. 
The simulation results are shown in Fig.~\ref{Joint_sum_rate_hierarhical_fig},
in which we compare the joint and single user decoding for both the closest BS and best channel user affiliation rules.
Fig.~\ref{Joint_sum_rate_hierarhical_fig}  demonstrates the performance improvement that incorporating virtual cells in the network provides, including the significant sum rate gain of fully centralized versus fully distributed optimization with joint decoding. It also shows that joint decoding can achieve significantly higher average system sum rate compared with single user decoding. However, single user decoding may yield higher sum rate in fully distributed setups in which ignoring out of cell interference affects the joint decoding scheme more severely since it depends on the covariance matrix of the interference and not just its diagonal.   

\section{Conclusion and Future Work}\label{sec:conclusion}
This work addresses the role of virtual cells in resource allocation for future CoMP cellular  networks. It addresses two design aspects of this optimization; namely, forming the virtual cells and allocating the communication resources in each virtual cell assuming cooperative decoding. We propose the use of  hierarchical clustering  in  forming the virtual cells so that changing the number of virtual cells only causes local changes and does not force a recalculation of all the virtual base-stations in the network. We also solved the uplink resource allocation problem optimally for cooperative decoding in each virtual cell.  Finally, we present numerical results for  these methods and discuss the merits of using virtual cells. We  note that other hierarchical  clustering algorithms can be considered in order to improve the overall network performance. Additionally, other models of cooperation in virtual cells can be considered as well.

\bibliographystyle{IEEEtran}



\end{document}